\documentstyle[12pt]{article}
\begin{document}
\begin{center}
{\bf \large Comment on "Weak localization in GaMnAs: evidence of
impurity band transport" by L. P. Rokhinson et. al. ( Phys. Rev.
B, 76, 161201 R; arXivCond-mat:0707.2416)}

{ N.V.Agrinskaya and V.I.Kozub}

{\it A.F.Ioffe Physico-Technical Institute, St.-Petersburg,
Russia}
\end{center}

In  the paper \cite{L} a presence of a small peak of negative
magnetoresistance at small temperatures and weak magnetic fields
(zero field peak) in the samples $Ga_{1-x}Mn_xAs$ was reported.
The authors attributed this peak to a manifestation of weak
localization within the impurity band. The important features of
the effects noted by the authors were the following:

"The height and shape of the zero-field peak is independent of the
orientation of magnetic field The overall shape of
magnetoresistance is very different for the two field
orientations, and exhibits a hysteretic behavior. The zero-field
peak, however, has no hysteresis and has a similar height and
width for both field orientations, which suggests that its origin
is not related to ferromagnetic ordering."

"The height of the peak then gradually increases as the
temperature decreases, and approaches 1-2$\%$ of the overall
resistivity at T=30 mK. The peak width also increases with
decreasing T, and almost saturates for $T <1$ K. The height and
shape of the zero-field peak is independent"

"Finally, we would like to point out several unusual features
observed in our experiments. Typically a negative
magnetoresistance in 3D disordered nonmagnetic conductors has
$B^2$ field dependence at low $B$, which smoothly evolves into
$B^{1/2}$ at higher $B$... In our samples, however, instead of
such gradual change of magnetoresistance with field we observe an
abrupt suppression of the effect. A related feature is the T
dependence of the width of the magnetoresistance peak ... one
expects that the peak will broaden with increasing temperature… In
our data, however, we observe just the opposite: The
magnetoresistance peak narrows as the temperature increases. "

According to the data, the peak disappeared at temperatures higher
than 3.4 K and the width of the peak at small temperatures was
around 30 mT.

We  would like to  emphasize that we observed earlier a similar
effect on completely different structures of GaAs/AlGaAs doped
quantum wells where the weak localization conductivity over
impurity band was realized \cite{ours}. While this conductivity
exhibited antilocalization positive magnetoresistance , the small
peak of negative magnetoresistance was observed at temperatures
lower than 3.4 K which was suppressed in magnetic fields higher
than 20 mT. We attributed such a behavior to a role of
superconducting indium leads with critical temperature $T_c$
around 3.4 K. We assumed that there is a tunnel barrier between
the contacts and  the sample. Thus, the Josephson tunneling was to
some extent suppressed while the single particle tunneling
suffered the superconducting gap. Correspondingly, the magnetic
field, suppressing the gap, decreased the contact resistance. The
saturation magnetic field corresponds to a critical magnetic field
of the superconductor which indeed saturates at low temperatures
while decreasing with increase of temperature which explains the
decrease of the peak width. In its turn, the peak height saturates
at temperatures where the single particle tunneling compares with
the Josephson tunneling which allowed us to estimate the tunneling
transparency which in our case was $\sim 10 \%$. Then, we also
related the manifestation of the effect to the delocalization
within the impurity band decreasing the resistance of the sample
and allowing to observe an effect of contact resistance.

Note that our measurements reported in \cite{ours} were made by
four-probe technique in the current-controlled regime which would
be expected to exclude the role of contacts. However the latter
suggestion literally holds only for the ideal case of point-like
contacts. In reality the contacts have a finite size and in our
case this size was only by a factor of 7 smaller than typical
in-plane size of the sample. It is important that the probes were
metallic and in any case and in any case their resistivity was
much smaller than the resistivity of the sample. If the contact
resistance between the voltage probes and the sample was large
(much larger than the resistance of the corresponding region of
the sample) the probes did not affect the current flow in the
sample and the probes acquire potentials roughly equal to the
potentials of the sample in the points corresponding to the
centers of the probes. However if the contact resistance becomes
smaller than the resistance of the surrounding region of the
sample, the probes act as local "shunts" for the sample affecting
the current distribution. In particular, the corresponding
correction of the potential within the sample at the distance $R$
from the probe can be estimated as $(Ed)d/R$ where $E$ is the
electric field, $d$ is the size of the contact. Thus the measured
voltage {\it appears to be sensitive to the contact resistance,
the sensitivity depends on the area of the probes}. In our case we
believe that the contact resistance in the superconducting state
of the probe is large enough to prevent the effect of the probes
on the current distribution within the sample. The magnetic field
suppresses this resistance leading to the "shunting" mentioned
above which reduces the measured potential drop, the relative
reduction is given as as $\sim d^2/L^2$ where $L$ is a distance
between the probe contacts in the direction of the current flow.
In our experiments the corresponding coefficient can be as large
as several percents which is in agreement to our data.

Note, that, as far as we  know, both in 2-terminal and 4-terminal
measurements the experimentalists often register features of weak
field magnetoresistance at temperatures around critical
temperature of the probes and believe that these features are in
some way related to superconducting transition in In probes.
However until now most of the experimentalists considered it as
parasitic effect. In contrast, in our paper we attracted attention
to this interesting phenomenon.



Another important factor noted in \cite{ours} is the sensitivity
of the effect observed to specific realization of the contacts. It
is explained, on the one hand, by a sensitivity of the contact
resistance to the tunneling barrier mentioned above, on the other
hand, by geometrical factor discussed at the previous paragraph.
In particular, the latter factor can explain the fact why the
effect of the superconducting leads was in Ref.\cite{Grbic}
observed in two-probe measurements and was not in four-probe
measurements while we observed this effect in four-probe
measurements.

We suspect that the effect observed in \cite{L}  may have a
similar nature since, as far as we know, the leads to the samples
GaAsMn are often fabricated from In. In particular, our model
seems to explain all the inconsistencies with weak localization
scenario noted by the authors of \cite{L} which were mentioned
above. We also note that, on the one hand, the thickness of the
sample in \cite{L} was expected to be smaller than the size of the
probes (which supports our estimates given above), on the second
hand, the sheet resistance of the sample in \cite{L} was less than
in our experiments which emphasize the role of the probe contacts.

\end{document}